\documentclass[aps,prd,showpacs,preprintnumbers,nofootinbib]{revtex4}
\usepackage[dvips]{graphicx}
\usepackage{bm,latexsym,amsmath,amssymb,amsfonts}
\newcommand*{\cM}{{\cal M}}
\newcommand*{\brane}{{\rm brane}}

\begin{document}
\title{Hybrid compactifications and brane gravity in six dimensions}

\author{Tsutomu Kobayashi}
\email{tsutomu"at"gravity.phys.waseda.ac.jp}
\author{Yu-ichi Takamizu}
\email{takamizu"at"gravity.phys.waseda.ac.jp}

\affiliation{
Department of Physics, Waseda University, Okubo 3-4-1, Shinjuku, Tokyo 169-8555, Japan
}

\begin{abstract}
We consider a six-dimensional axisymmetric Einstein-Maxwell model of warped braneworlds.
The bulk is bounded by two branes, one of which is a conical 3-brane
and the other is a 4-brane wrapped around the axis of symmetry.
The latter brane is assumed to be our universe.
If the tension of the 3-brane is fine-tuned,
it folds the internal two-dimensional space in a narrow cone,
making sufficiently small the Kaluza-Klein circle of the 4-brane.
An arbitrary energy-momentum tensor
can be accommodated on this ring-like 4-brane.
We study linear perturbations sourced by matter on the brane, and show that
weak gravity is apparently described by a four-dimensional scalar-tensor theory.
The extra scalar degree of freedom can be interpreted as
the fluctuation of the internal space volume
(or that of the circumference of the ring), the effect of which turns out to be
suppressed at long distances. Consequently, four-dimensional Einstein gravity is
reproduced on the brane.
We point out that as in the Randall-Sundrum model, the brane bending mode
is crucial for recovering the four-dimensional tensor structure in this setup.
\end{abstract}

\preprint{WU-AP/271/07}

\maketitle

\section{Introduction}

Probably one of the most interesting recent developments
in particle physics and cosmology has been the idea of braneworlds.
Models with extra dimensions are motivated theoretically,
as in superstring theory,
which is a very promising approach
to unification, requiring ten spacetime dimensions.
Braneworld scenarios are further motivated by their phenomenologically interesting aspects.
Among them are the possible effect of having the fundamental scale as low as the weak scale
and some modification of the gravity law on submillimeter scales~\cite{ADD, AADD, RS1, RS2},
both of which are accessible by experiments.
A string realization of the scenario of~\cite{ADD, AADD} is found in~\cite{ck}.
So far five-dimensional (5D) Randall-Sundrum-type braneworlds~\cite{RS1, RS2} have been
the most extensively studied examples,
whereas more recently there has been growing interest
in six- or higher dimensional models~\cite{6d_review, vin}.

In the present paper we will be focusing on 6D braneworlds with Maxwell fields.
Since two extra dimensions are enough to admit flux-stabilized compactifications
while keeping the setup as simple as possible,
such brane models allow us to explore some of the interesting features
which would be less easily addressed in more string theoretical settings.\footnote{
It should be noted that
the 6D brane models with football-shaped extra dimensions
have attracted much attention due to their potential mechanism
for resolving the cosmological constant problem~\cite{cc, cc2}.
For a comprehensive review of the cosmological constant and dark energy
in braneworlds, see Ref.~\cite{koyama_cc}.}
Perhaps the simplest exact solution of this type of warped braneworlds
has been constructed in~\cite{Mukohyama}, and
subsequent work has addressed the stability issue of
this model~\cite{Yoshiguchi, Sendouda, Kinoshita} (see also~\cite{deRham:2005ci}).
Braneworlds in 6D supergravity
have also been much investigated: solutions with 4D maximal
symmetry~\cite{Gibbons,A_et_al,Burgess1,dS_Tolley}
or time-dependent dynamics~\cite{Scaling, Kobayashi_Minamitsuji, Copeland_Seto} have been found,
and the behavior of perturbations has been studied in~\cite{Lee-P, Kick, Parameswaran}.
Codimension two branes are often considered in the above approaches,
and they are unfortunately associated with the problem of the localization of matter. Namely,
a strict codimension two defect does not allow for arbitrary energy-momentum tensor
localized on it~\cite{putmat}.
(Branes with codimension higher than two make the situation worse
if one attempts to construct a brane model in seven or higher dimensions
while taking seriously into account self-gravity of the branes,
as they develop spacetime singularities.)
Gravitational aspects of such higher dimensional braneworlds
have not been explored thoroughly yet because of this fact.
The hybrid Kaluza-Klein / Randall-Sundrum construction of~\cite{bolts}
evades this problem by
assuming that our universe is a 4-brane in six dimensions,
with one of the spatial directions compactified on a circle
(see~\cite{hbss} for a supergravity generalization).
Refs.~\cite{Peloso1, Peloso2, PPZ, extended, uvcaps}
also exploit essentially
the same idea to resolve codimension two singularities (see also~\cite{Yamauchi}).

The specific model we consider in this paper
is most closely similar to that of~\cite{bolts},
but not exactly the same.
In~\cite{bolts} the bulk with axisymmetry closes regularly at the point
where the axial Killing vector vanishes.
In contrast, ours does not, permitting a conical singularity there,
corresponding to a tensional 3-brane.
The 3-brane can fold the internal 2D space in a narrow cone, yielding
a small Kaluza-Klein circle of the 4-brane wrapped around the symmetry axis.
(For this idea we are indebted to~\cite{sliver}.) 
The analysis using a massless minimally coupled scalar field
has shown that the static scalar potential has a long-distance behavior
proportional to $-|\mathbf{x}-\mathbf{x}'|^{-1}$~\cite{bolts},
from which one may expect that standard Newtonian gravity is reproduced on the brane.
However, Ref.~\cite{bolts} has not given a complete analysis
of gravitational perturbations, and the story will be more complicated.
To study in more detail the behavior of weak gravity sourced by matter in the braneworld,
we provide a
rigorous treatment of metric and matter perturbations in this paper.
We use the technique of~\cite{G-T},
which was originally developed for studying linear perturbations
in the Randall-Sundrum model and was developed by~\cite{Peloso1, extended}
in the context of 6D brane models.


The plan of the paper is as follows.
In the next section we present the description of our braneworld model.
Then in section~\ref{sec:pert} we perform a linear analysis
of gravity, showing in detail the mechanism for recovering 4D Einstein gravity.
Section~\ref{sec:final} is devoted to discussion.

\section{The model}

\subsection{Bulk geometry}

Our 6D bulk is described by the Einstein-Maxwell action
\begin{eqnarray}
S=\int d^6x\sqrt{-g}
\left[\frac{1}{2\kappa^2}\left(R-2\Lambda_6\right)-\frac{1}{4}F_{MN}F^{MN}\right],
\end{eqnarray}
where $F_{MN}:=\partial_MA_N-\partial_NA_M$ is the field strength of the $U(1)$ gauge field.
In our setup
the bulk cosmological constant may be positive or negative or zero,
and so we write
\begin{eqnarray}
\Lambda_6=\epsilon \frac{10}{\ell^2},\quad \epsilon= \pm 1,\;0.
\end{eqnarray}
The 6D field equations derived from the above action are
\begin{eqnarray}
\text{(Einstein)}:&&\;
R_{MN}-\frac{1}{2}g_{MN}R=-\epsilon\frac{10}{\ell^2} g_{MN}
+\kappa^2\left(F_{ML}F_N^{\;\;L}-\frac{1}{4}g_{MN}F^2\right),
\label{einstein}
\\
\text{(Maxwell)}:&&\;
\partial_M\left(\sqrt{-g}F^{MN}\right) =0. \label{maxwelleqs}
\end{eqnarray}

The field equations~(\ref{einstein}) and~(\ref{maxwelleqs}) admit
the following bulk solution~\cite{Membrane, bolts}:
\begin{eqnarray}
g_{MN}dx^Mdx^N=\xi^2\eta_{\mu\nu}dx^{\mu}dx^{\nu}+\ell^2
\left[\frac{d\xi^2}{f(\xi)}+\beta^2f(\xi)d\theta^2\right],
\end{eqnarray}
where
\begin{eqnarray}
f(\xi):=-\epsilon\xi^2+\frac{\mu}{\xi^3}-\frac{q^2}{\xi^6}
\end{eqnarray}
and $\beta$ is an arbitrary constant at this stage.
We assume here that $\mu$ is positive.
Only the $(\xi\theta)$ component of the field strength is nonvanishing; it is given by
\begin{eqnarray}
F_{\xi\theta}=2\sqrt{3}\frac{\beta \ell}{\kappa}\frac{q}{\xi^4}.
\end{eqnarray}

Let $\xi_0$ be the positive zero of $f(\xi)$.
We consider the region in which $\xi\geq \xi_0$ and $f(\xi)\geq 0$.
More specifically,
$\xi_0$ is the largest positive zero of $f(\xi_0)$ for $\epsilon=-1$.
For $\epsilon=0$, we have $\xi_0=(q^2/\mu)^{1/3}$.
In the $\epsilon=1$ case, $\xi_0$ is the second largest positive zero, and
we consider the region $\xi_0\leq\xi<\xi_1$, with $\xi_1$ being the largest zero.

Since $F_{\xi\theta}=A_{\theta}'$, where a prime stands for a derivative with respect to $\xi$,
we have
\begin{eqnarray}
A_{\theta}=-\frac{2q}{\sqrt{3}}\frac{\beta\ell}{\kappa}\left(\frac{1}{\xi^3}
-\frac{1}{\xi_0^3}\right),
\end{eqnarray}
where the integration constant has been chosen so that $A_{\theta}(\xi_0)=0$.

We assume that $\theta$ has period $2\pi$.
Accordingly, we have a deficit angle $\delta=2\pi\left[1-\beta f'(\xi_0)/2\right]$,
corresponding to a conical 3-brane placed at $\xi=\xi_0$ with tension
\begin{eqnarray}
\kappa^2\sigma=2\pi\left[1- \frac{\beta f'(\xi_0)}{2}\right].
\end{eqnarray}
As in~\cite{bolts},
one may impose $\beta=2/f'(\xi_0)$,
leading to the regular geometry without a 3-brane.
In the present paper, however, we do not do so and allow for a conical deficit.

\subsection{Adding a 4-brane}\label{add4brane}

We follow the construction of~\cite{bolts}
and add a ring-like 4-brane at a point $\xi_*>\xi_0$,
which is assumed to be our universe.
The brane action is given by
\begin{eqnarray}
S_{\brane}=\int d^5x\sqrt{-\gamma}\left(- \lambda+{\cal L}_m\right),
\end{eqnarray}
where $\lambda$ is the tension of the 4-brane and
${\cal L}_m$ is the matter Lagrangian.
We denote by $\gamma_{ab}$ the induced metric on the brane. 
Let ${\cal M}$ be the spacetime in which $\xi$ ranges from $\xi_0$ to $\xi_*$.
We impose $\mathbb{Z}_2$ symmetry about $\xi_*$, and
glue $\cM$ and a copy of $\cM$ together at $\xi=\xi_*$.
In so doing we assume that the metric and $F_{MN}$ are continuous across the
brane.\footnote{We impose the same boundary condition
as in~\cite{bolts} for the Maxwell field.
This is different from~\cite{Peloso1, Peloso2, PPZ, extended}, in which
$F_{MN}$ is discontinuous at the 4-brane due to
the St\"{u}ckelberg term included in the brane action.}
The first derivative of the metric is subject to the Israel conditions
\begin{eqnarray}
K_{ab}-K\gamma_{ab}=\frac{\kappa^2}{2}\lambda\gamma_{ab}-\frac{\kappa^2}{2}T_{ab},
\end{eqnarray}
where $K_{ab}:=\gamma_a^{\;c}\gamma_b^{\;d}\nabla_{(c}n_{d)}$
is the extrinsic curvature on the brane and
$T_{ab}$ is the energy-momentum tensor of brane matter.
The unit normal to the brane $n_a$ is defined as
pointing inside $\cM$.

We now consider a vacuum brane (i.e., $T_{ab}=0$). In this case the Israel conditions read
\begin{eqnarray}
(\mu\nu):
&&\quad
\frac{\sqrt{f_*}}{\ell}\left(\frac{f'_*}{2f_*}+\frac{3}{\xi_*}\right)=\frac{\kappa^2\lambda}{2},
\label{Jmn}
\\
(\theta\theta):&&\quad
\frac{\sqrt{f_*}}{\ell} \frac{4}{\xi_*} =\frac{\kappa^2\lambda}{2},
\label{Jtt}
\end{eqnarray}
where various quantities with $*$ are evaluated at $\xi=\xi_*$.
Eliminating $\lambda$ we find $\xi_*f'_*=2f_*$, which determines the brane position as
\begin{eqnarray}
\xi_*=2\left(\frac{q^2}{5\mu}\right)^{1/3}. \label{xi_*=}
\end{eqnarray}
The two conditions~(\ref{Jmn}) and~(\ref{Jtt}) completely fix
the position of the brane.
This is in contrast to the Randall-Sundrum model~\cite{RS1, RS2}, in which
the brane positions are arbitrary.

\begin{figure}[t]
  \begin{center}
    \includegraphics[keepaspectratio=true,height=30mm]{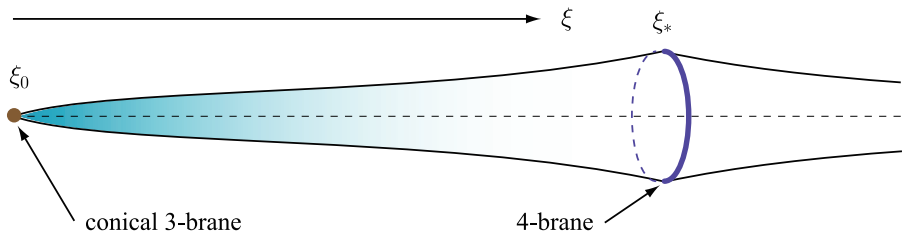}
  \end{center}
  \caption{The sliver-shaped bulk.}%
  \label{fig:braneworld.eps}
\end{figure}

Since our brane model includes one Kaluza-Klein direction,
we must impose that the circumference of the ring,
\begin{eqnarray}
{\cal C}=2\pi\beta\ell\sqrt{f_*},
\end{eqnarray}
is not too large (say ${\cal C}\lesssim 10^{-16}$\;cm),
whereas if
the scale of the ``braneworld compactification'' is as large as $\ell \sim 10^{-2}$\;cm
it will be particularly interesting.
Clearly, this can be achieved by requiring $\beta\sqrt{f_*}\ll 1$.
In other words, if the tension of the conical brane is
fine-tuned to be very close to the critical value, $\kappa^2\sigma \simeq 2\pi$,
the bulk will look like a narrow sliver with a small Kaluza-Klein circle
(figure~\ref{fig:braneworld.eps}).
The required fine-tuning is\footnote{In the case of $\epsilon=0$,
$\ell$ in Eq.~(\ref{fine-tune}) should be replaced by $\ell_0$ defined below.}
\begin{eqnarray}
1-\frac{\kappa^2\sigma}{2\pi}\sim \frac{{\cal C}}{\ell}.\label{fine-tune}
\end{eqnarray}
Note in passing that in this setup both branes have positive tension.

If one wishes to avoid the conical singularity at $\xi=\xi_0$,
the regularization procedure as in~\cite{Peloso1, Peloso2, PPZ, extended} will be helpful.
After replacing the conical 3-brane by an extended 4-brane
in an appropriate manner,
one can still attain a narrow cone-shaped geometry.
However, for clarity we will keep using the conical brane to set the boundary of the system.

\paragraph{Reparameterization}

Using Eq.~(\ref{xi_*=}) and the condition $f(\xi_0)=0$,
we can express the parameters $\mu$ and $q^2$ in terms of $\xi_0$ and $\xi_*$:
\begin{eqnarray}
\mu=-\epsilon\frac{8\xi_0^5}{5\alpha^3-8},
\qquad
q^2=-\epsilon\frac{5\alpha^3\xi_0^8}{5\alpha^3-8},
\qquad\text{where}\qquad \alpha:=\frac{\xi_*}{\xi_0}.
\label{muq^2}
\end{eqnarray}
Note that the above expression is valid only for $\epsilon\neq0$.
Introducing the new coordinate $z:=\xi/\xi_0$, we write $f=\xi_0^2\overline{ f}(z)$, where
\begin{eqnarray}
\overline{f}(z):=-\epsilon\left(
z^2+\frac{8}{5\alpha^3-8}\frac{1}{z^3}
-\frac{5\alpha^3}{5\alpha^3-8}\frac{1}{z^6}
\right).\label{bar:f}
\end{eqnarray}
The background solution apparently depends on $\xi_0$, but
it can be eliminated by performing an appropriate coordinate rescaling.
Thus, it turns out that the background configuration in the $\epsilon\neq0$ models
is characterized by two parameters, $\alpha$ and the 3-brane tension $\sigma$.
The expression~(\ref{bar:f}) is sometimes convenient as
it includes only a single parameter $\alpha$.

From Eq.~(\ref{muq^2}) we see that
$(1<) \;\alpha^3 <8/5$ for $\epsilon=+1$ and
$\alpha^3 >8/5$ for $\epsilon=-1$.
If $\alpha$ is very close to $2/5^{1/3}$, we have
a large circumference, ${\cal C}\propto |\alpha-2/5^{1/3}|^{-1/2}$.
Large $\alpha$ also tends to give a large Kaluza-Klein radius, ${\cal C}\propto \alpha$
(i.e., the conical brane fails to reduce the circumference of the 4-brane
placed too far from it).
Therefore, in what follows we will assume $\alpha\sim O(1)$ but not too close to $2/5^{1/3}$.

The special case with $\epsilon=0$ ($\alpha^3=8/5$) should be considered separately.
Since the 6D cosmological constant vanishes, the typical compactification scale
is given solely by the Maxwell field: $\kappa^2 F^2=(24/\ell^2_0) z^{-8}\sim 1/\ell^2_0$, where
\begin{eqnarray}
\ell_0 := \frac{\xi_0^4\ell}{q}.
\end{eqnarray}
This is an integration constant of the solution but not a parameter included in the Lagrangian.
Therefore, though $\alpha$ is fixed, the background solution
still has two parameters: $\ell_0$ and $\sigma$.
Note that the metric of the 2D internal space can in fact be written as
\begin{eqnarray*}
\ell^2\frac{d\xi^2}{f(\xi)}+\cdots = \ell_0^2\frac{dz^2}{\overline{f}(z)}+\cdots,
\qquad\text{with}\qquad
\overline{f}(z):=z^{-3}-z^{-6}.
\end{eqnarray*}

\section{Linear perturbations}\label{sec:pert}

Let us now analyze linear perturbations on the brane model
described in the previous section.
We are interested in a length scale much larger than the circumference of the ring,
and hence we focus on perturbations homogeneous in the $\theta$-direction.

\subsection{Perturbation equations and boundary conditions}

Linear perturbations are split into scalar, vector, and tensor modes under the Lorentz group
in the external spacetime.
Since they do not mix with one another in the 6D field equations,
equations of motion for each mode can be studied separately.
Here let us consider scalar and tensor perturbations.
(Vector modes are of no particular interest.)
The perturbed metric in an arbitrary gauge can be written as
\begin{eqnarray}
&&\left(g_{MN}+\delta g_{MN}\right)dx^Mdx^N =
\xi^2\left[(1+2\Psi)\eta_{\mu\nu}+2E_{,\mu\nu}
+h_{\mu\nu}\right]dx^{\mu}dx^{\nu}
+2B_{,\mu}d\xi dx^{\mu}
\nonumber\\
&&\quad
+2D_{,\mu}d\theta dx^{\mu}
+\ell^2\left[\left(1+2\Xi\right)\frac{d\xi^2}{f}+2\beta^2fTd\xi d\theta
+(1-2\Omega-6\Psi)\beta^2 fd\theta^2\right],
\end{eqnarray}
and the perturbed gauge field is
\begin{eqnarray}
\delta A_M=\left( \delta A_{,\mu}, \delta A_{\xi}, \delta A_{\theta}\right).
\end{eqnarray}

For the transverse and traceless tensor perturbation, $h_{\mu\nu}$,
the Einstein equations simply give~\cite{Yoshiguchi, extended}
\begin{eqnarray}
\left(\xi^4 f h_{\mu\nu}'\right)'+\xi^2\ell^2\Box h_{\mu\nu}=0,
\end{eqnarray}
where $\Box:=\eta^{\mu\nu}\partial_{\mu}\partial_{\nu}$.

For the scalar perturbations, we begin with fixing the gauge freedom and reduce
the number of modes that we consider.
To study the 6D field equations it is convenient to employ
the gauge defined by $E=B=T=0$, which we denote as
the 6D longitudinal gauge (see Appendix~\ref{App:gauge}).
The $(\xi\theta)$ component of the Einstein equations implies
\begin{eqnarray}
\left(f^{-1}\Box D\right)'=0,
\end{eqnarray}
and so $D$ can be set to be zero by using the residual gauge freedom
$\theta\to\theta+\delta\theta(x)$~\cite{Yoshiguchi}.
Then the $(\mu\theta)$ component of the Einstein equations leads to
$\delta F_{\mu\xi}=\left(\delta A_{\xi}-\delta A\right)_{,\mu}=0$.

The $(\mu\nu)$, $(\xi\xi)$, and $(\theta\theta)$ components
of the Einstein equations are combined to give~\cite{Yoshiguchi, extended}
\begin{eqnarray}
\Omega''+2\left(\frac{f'}{f}+\frac{5}{\xi}\right)\Omega'-\epsilon\frac{40}{f}(\Omega+\Psi)+
\frac{\ell^2}{\xi^2f}\Box\Omega&=&0, \label{master1}
\\
\Psi''+\frac{4}{\xi}\Psi'+\frac{\ell^2}{2\xi^2 f}\Box(\Omega+2\Psi)&=&0. \label{master2}
\end{eqnarray}
The remaining variables are obtained from
\begin{eqnarray}
\Xi&=&\Psi+\Omega, \label{Xi=}
\\
\delta A_{\theta}&=&\frac{\beta\ell\xi^3}{2\sqrt{3}\kappa q}\left[
f\left(\xi\Omega'+2\Omega\right)+\xi f'(\Omega+2\Psi)
\right], \label{dA=}
\end{eqnarray}
which are the traceless part and ($\mu\xi$) component of the Einstein equations, respectively.
The perturbed Maxwell equations can be derived from
the above Einstein equations.

We now proceed to discuss boundary conditions.
At the point where the geometry pinches off, $\xi=\xi_0$,
we impose some regularity conditions on the perturbations.
For the tensor mode, we require that both $h_{\mu\nu}$ and $h_{\mu\nu}'$
are regular at $\xi=\xi_0$.
The regularity conditions for the scalar modes are~\cite{Yoshiguchi, Sendouda}
\begin{eqnarray}
f\Omega|_{\xi_0}&=&0, \label{reg1}
\\
(f\Omega)'+2f'\Psi|_{\xi_0}&=&0. \label{reg2}
\end{eqnarray}
The above boundary conditions do not include information
on the conical 3-brane (i.e., the brane tension $\sigma$).
This means that
the dynamics of axisymmetric perturbations does not depend on
how the bulk closes at $\xi_0$.
The conical brane is introduced for the purpose
of reducing the size of the Kaluza-Klein circle.

The perturbed field strength,
$\overline{\delta F}_{\xi\theta}={\overline{\delta A}_{\theta}}'$ and
$\overline{\delta F}_{\mu\theta}=\overline{\delta A}_{\theta,\mu}$,
must be continuous at $\xi=\xi_*$, where
we denote by a bar the perturbations in the Gaussian-normal gauge (see Appendix~\ref{App:gauge}).
Since we are assuming the $\mathbb{Z}_2$ symmetry across the ring,
it is required that $\overline{\delta A}_{\theta*}=0$,
leading to the condition
\begin{eqnarray}
\delta A_{\theta*}+A_{\theta*}'\zeta=0, \label{continuity:maxwell}
\end{eqnarray}
where
the equation is written in terms of the 6D longitudinal gauge perturbations
and hence includes the brane bending mode $\zeta=\zeta(x)$.
(In the 6D longitudinal gauge, the location of the brane is perturbed
in general: $\xi_*\to\xi_*+\zeta(x)$.)

The Israel conditions at the ring are given by
\begin{eqnarray}
\left.\frac{\sqrt{f}}{\ell}\left[\frac{\ell^2}{f}\left(\Box\zeta\eta_{\mu\nu}-\zeta_{,\mu\nu}\right)
+\frac{\xi^2}{2}h_{\mu\nu}'\right]\right|_{\xi_*}
=\frac{\kappa^2}{2}T_{\mu\nu},\label{Is:munu}
\end{eqnarray}
and
\begin{eqnarray}
\left.\frac{\sqrt{f}}{\ell}\left(\frac{\ell^2}{\xi^2f}\Box\zeta-4\Psi'+\frac{4}{\xi}\Xi\right)
\right|_{\xi_*}
=\frac{\kappa^2}{2}T_{\theta}^{\;\theta},\label{Is:thth}
\end{eqnarray}
where we used Eq.~(\ref{continuity:maxwell}) to simplify the first equation.

\subsection{Zero-mode truncation and linearized gravity}

Following~\cite{G-T} (and~\cite{Peloso1, extended}),
we now investigate the long-distance behavior of weak gravity on the 4-brane.

The Israel condition~(\ref{Is:munu}) can be rearranged to give
\begin{eqnarray}
\left.h_{\mu\nu}'\right|_{\xi_*}=\frac{\ell}{\xi^2_*\sqrt{f_*}}\kappa^2
\left(T_{\mu\nu}-\frac{1}{3}T_{\lambda}^{\;\lambda}\gamma_{\mu\nu}\right)
+\frac{2\ell^2}{\xi^2_*f_*} \zeta_{,\mu\nu}=: \frac{{\cal S}_{\mu\nu}}{2\xi_*^4f_*},
\label{Is:h_prime=}
\end{eqnarray}
where we used the trace of~(\ref{Is:munu}):
\begin{eqnarray}
\kappa^2 T_{\mu}^{~\mu}
=\frac{6\ell }{\xi_*^2 \sqrt{f_*}}\Box\zeta.\label{trace:Is}
\end{eqnarray}
Using Eq.~(\ref{Is:h_prime=}) we can put the bulk equation of motion and
the boundary condition into a single equation with a source term:
\begin{eqnarray}
{\cal O}h_{\mu\nu}:=\left(\xi^4 fh_{\mu\nu}'\right)'+\xi^2\ell^2\Box h_{\mu\nu}
=-{\cal S}_{\mu\nu}\delta(\xi-\xi_*).\label{eom+bc}
\end{eqnarray}
We use the standard Green function method
to solve Eq.~(\ref{eom+bc}). The Green function
satisfies
${\cal O}G_R(x,\xi;x', \xi')=\delta^{(4)}(x-x')\delta(\xi-\xi')$,
in terms of which we have
\begin{eqnarray}
h_{\mu\nu}(x, \xi)=-\int d^4x'G_R(x,\xi; x', \xi_*){\cal S}_{\mu\nu}.
\end{eqnarray}
The Green function is explicitly given by
\begin{eqnarray}
G_R(x,\xi ; x', \xi')=-\int \frac{d^4k}{(2\pi)^4}e^{ik\cdot(x-x')}\sum_i
\frac{u_i(\xi)u_i(\xi')}{m_i^2+\mathbf{k}^2-(\omega+i\epsilon)^2},
\end{eqnarray}
where $u_i(\xi)$ are a complete set of eigenfunctions of
\begin{eqnarray}
\left(\xi^4 f u_i'\right)'=-\xi^2\ell^2m_i^2u_i. \label{mode-eq}
\end{eqnarray}
The eigenfunctions are normalized according to
\begin{eqnarray}
2\ell^2\int^{\xi_*}_{\xi_0} \xi^2u_i u_j d\xi=\delta_{ij}.\label{norm}
\end{eqnarray}

We are mainly interested in the long-range gravity on the brane and hence
the zero-mode solution of~(\ref{mode-eq}) is the most important.
Setting $m^2_0=0$ and integrating once, we obtain $u_0'=\xi^{-4}f^{-1}U$,
where $U$ is an integration constant. However, 
from the regularity condition at $\xi=\xi_0$ we must impose $U=0$.
Therefore, the zero-mode solution is given by
$u_0=L^{-1}=$ constant.
The normalization is determined by Eq.~(\ref{norm}) as
\begin{eqnarray}
L=\ell\sqrt{\frac{2}{3}(\xi_*^3-\xi_0^3)}.
\end{eqnarray}
The zero-mode truncation of the Green function~\cite{G-T} leads to
\begin{eqnarray}
h_{\mu\nu}\approx-\frac{1}{L^2}\Box^{-1} {\cal S}_{\mu\nu}.\label{truncated_sol}
\end{eqnarray}

Now we would like to compute the Ricci tensor $R_{\mu\nu}^{(4)}$ of the 4D metric
$\overline{g}_{\mu\nu}=\xi_*^2[ (1+2\overline{\Psi}_* )\eta_{\mu\nu}+h_{\mu\nu} ]$.
Here $\overline{\Psi}_*$ is the metric perturbation in the Gaussian-normal gauge,
which is related to the longitudinal gauge quantities via Eq.~(\ref{induced_pert}).
Following~\cite{Peloso1, extended} we write
\begin{eqnarray}
R^{(4)}_{\mu\nu}&=&-\frac{1}{2}\Box h_{\mu\nu}-2\overline{\Psi}_{*,\mu\nu}
-\Box\overline{\Psi}_*\eta_{\mu\nu}
\nonumber\\
&=&
-\frac{1}{2}\Box h_{\mu\nu}-\frac{2\xi_*^2\ell^2}{L^2}\zeta_{,\mu\nu}
-\frac{\ell^2}{L^2}\gamma_{\mu\nu}\Box\zeta
-\left(2\partial_{\mu}\partial_{\nu}+\eta_{\mu\nu}\Box\right)\Upsilon,\label{4d-ricci}
\end{eqnarray}
where we defined
\begin{eqnarray}
\Upsilon:= \overline{\Psi}_*-\frac{\ell^2}{L^2}\xi_*^2\zeta.
\end{eqnarray}
Using Eqs.~(\ref{truncated_sol}) and~(\ref{trace:Is}), we find
\begin{eqnarray}
R^{(4)}_{\mu\nu}
 \approx \kappa_4^2 \left(\overline{T}_{\mu\nu}-\frac{1}{2}\overline{T}_{\lambda}^{\;\lambda}
\gamma_{\mu\nu}\right)
-\left(2\partial_{\mu}\partial_{\nu}+\eta_{\mu\nu}\Box\right)\Upsilon,
\label{ricci==}
\end{eqnarray}
where $\overline{T}_{ab}:={\cal C} T_{ab}$ is the
energy-momentum tensor integrated along the $\theta$-direction,
and we defined the 4D Newton constant as
\begin{eqnarray}
\kappa_4^2:=\frac{\xi_*^2\kappa^2}{2\pi L^2\beta}.\label{newton_const}
\end{eqnarray}
Thus, we see that the first three terms in~(\ref{4d-ricci})
help to recover a 4D gravitational theory.
However, brane gravity looks different from Einstein gravity at this stage
because of the additional scalar degree of freedom encoded in $\Upsilon$.
It should be stressed here that
{\em the brane bending mode is crucial for reproducing the 4D tensor structure.}
The role of the brane bending here is the same as
that of the Randall-Sundrum braneworld~\cite{G-T},
and it has been shown that the same mechanism works
in a slightly different setup of 6D braneworlds~\cite{Peloso1, extended}.

Let us evaluate the effect of $\Upsilon$.
For this purpose it is a good approximation to set $\ell^2\Box \approx 0$
in Eqs.~(\ref{master1}) and~(\ref{master2}), picking up zero-mode contributions.
For $\ell^2\Box=0$ we have the following exact solutions:
\begin{eqnarray}
\Omega_0&=&\frac{1}{f}\left[
\epsilon\left(c_1\xi^2+\frac{c_2}{\xi}\right)+\frac{c_3}{\xi^3}+\frac{c_4}{\xi^6}\right],
\\
\Psi_0&=&c_1+\frac{c_2}{4\xi^3},
\end{eqnarray}
where integration constants $c_1(x), \cdots$\;etc. are to be determined by the
boundary conditions.
In the absence of matter excitations, we can easily see that
no scalar modes are present, $c_1=c_2=c_3=c_4=0$. However,
in general cases with $T_{ab}\neq0$ we have nonzero integration constants.
From the regularity conditions~(\ref{reg1}) and~(\ref{reg2}),
one can express $c_3$ and $c_4$ in terms of $c_1$ and $c_2$.
Then, Eq.~(\ref{continuity:maxwell}),
with the aid of Eq.~(\ref{dA=}),
allows one to write $\zeta$ in terms of $c_1$ and $c_2$. For $\epsilon\neq0$ we find
\begin{eqnarray}
\Upsilon =\frac{(\xi_*^3-\xi_0^3)(\xi_*^3+8\xi_0^3)}{72\xi_*^3\xi_0^6}\hat c(x),
\end{eqnarray}
where $\hat c:=8\xi_0^3c_1-c_2$.
Similarly, it follows that
\begin{eqnarray}
\Psi_*'-\frac{1}{\xi_*}\Xi_*= \frac{5\xi_0^2(\xi_*^3-\xi_0^3)^2}
{3\xi_*^4(5\xi_*^8-8\xi_*^5\xi_0^3+3\xi_0^8)}\hat c(x).
\end{eqnarray}
Using the Israel conditions~(\ref{Is:thth}) and~(\ref{trace:Is}), we finally arrive at
\begin{eqnarray}
\Upsilon= {\cal F}(\alpha)\ell^2  \;\kappa_4^2
\left(\frac{1}{3}\overline{T}_{\lambda}^{\;\lambda}
-\overline{T}_{\theta}^{\;\theta}\right),
\label{estimate_Upsilon}
\end{eqnarray}
where
\begin{eqnarray}
{\cal F}(\alpha):=-\frac{\epsilon}{1440}\alpha^2(5\alpha^3-8)(\alpha^3+8).
\end{eqnarray}
Eqs.~(\ref{ricci==}) and~(\ref{estimate_Upsilon}) imply that
the effect of $\Upsilon$ is suppressed on scales much greater than $\sqrt{{\cal F}}\ell$.
For $\alpha \sim O (1)$, the coefficient $\sqrt{{\cal F}}$ is not large,
so that the critical scale may be given by $\ell$.
The critical scale becomes large for $\alpha\gg 1$, but
this is not the case we are considering.

In the $\epsilon=0$ case, a straightforward computation similarly shows that $\Upsilon=\hat c/20\xi_0^3$
and $\Psi'-\Xi/\xi_*=5^{1/3}\hat c/16\xi_0^4$, leading to
\begin{eqnarray}
\Upsilon= \frac{16}{3\cdot 5^{8/3}}\ell^2_0\; \kappa_4^2\left(\frac{1}{3}
\overline{T}_{\lambda}^{\;\lambda}
-\overline{T}_{\theta}^{\;\theta}\right).
\end{eqnarray}
Therefore, in this case the effect of $\Upsilon$ is negligible on scales much greater than
$\ell_0$.

To illustrate the geometrical interpretation of the scalar mode $\Upsilon$,
we compute the perturbations of the internal space volume and
the circumference of the brane~\cite{Peloso1, extended},
\begin{eqnarray}
\delta{\cal V}=4\pi\ell^2 \beta\left(\zeta-2\int^{\xi_*}_{\xi_0}\Psi d\xi\right),
\qquad
\delta{\cal C}=2\pi\ell\beta\sqrt{f_*}\left(\frac{\zeta}{\xi_*}-\Omega_*-3\Psi_*\right).
\end{eqnarray}
It then turns out that
\begin{eqnarray}
\delta{\cal V}\propto \delta{\cal C}\propto \hat c.
\end{eqnarray}
Namely, 
$\Upsilon \;(\propto\hat c)$ can be interpreted as
the perturbations of the internal space volume and the circumference of the ring.
It is reasonable that
standard 4D gravity is recovered when
the matter fields on the brane do not perturb the internal space much.

In the present setup we are imposing the $\mathbb{Z}_2$ symmetry
and continuity of the $U(1)$ field strength at the ring.
These boundary conditions are different from those in~\cite{Peloso1, extended},
in which $F_{MN}$ has a jump at the ring and no $\mathbb{Z}_2$ symmetry
is assumed there.
Nevertheless, one notices that what happens here for recovering standard 4D gravity
is quite similar to what occurs in~\cite{Peloso1, extended}.


\subsection{Kaluza-Klein tensor modes}

So far we have seen that the zero-mode sector of perturbations
can reproduce standard 4D gravity on the brane.
Basically, the effect of discrete Kaluza-Klein modes are Yukawa-suppressed, and hence
we can safely neglect these massive modes at long distances.
In this subsection, we compute the mass spectrum of the Kaluza-Klein modes
for completeness.\footnote{The absence of tachyonic modes can be shown
as follows. From Eq.~(\ref{mode-eq-z}) we see that
\begin{eqnarray*}
\nu_i^2\int_1^{\alpha}z^2u_i^2dz=-\int_1^{\alpha}\frac{d}{dz}\left(z^4\overline{f} \frac{du_i}{dz}u_i\right)dz
+\int_1^{\alpha}z^4\overline{f}\left(\frac{du_i}{dz}\right)^2dz.
\end{eqnarray*}
The first term vanishes because $\overline{f}(1)=0$ and $du_i/dz|_{z=\alpha}=0$.
Thus, we have $\nu_i^2\geq0$.
}

To do so we rewrite Eq.~(\ref{mode-eq}) in terms of
$z$ and $\overline{f}(z)$ defined in section~\ref{add4brane},
so that we would like to solve
\begin{eqnarray}
\frac{d}{dz}\left[z^4\overline{f}(z)\frac{du_i}{dz} \right]+\nu_i^2z^2 u_i=0,
\qquad
\nu_i^2:=\frac{m_i^2\ell^2}{\xi_0^2},
\label{mode-eq-z}
\end{eqnarray}
supplemented with the boundary conditions
\begin{eqnarray}
\left.\frac{d\overline{f}}{dz}\frac{du_i}{dz}+\nu_i^2u_i\right|_{z=1}=0,
\qquad
\left.\frac{du_i}{dz}\right|_{z=\alpha}=0.\label{bc-z}
\end{eqnarray}
For $\epsilon=0$ we replace $\ell^2$ in $\nu_i^2$ by $\ell_0^2$.

In the case of $\epsilon=0$ we have analytic solutions for the Kaluza-Klein mode functions.
Using $y:=(z^3-1)^{1/2}$, Eq.~(\ref{mode-eq-z}) can be rewritten
in the form of the Bessel equation:
\begin{eqnarray}
\frac{1}{y}\frac{d}{dy}\left(y\frac{du_i}{dy} \right)+\frac{4}{9}\nu^2_i u_i=0.
\end{eqnarray}
The solution regular at $z=1\;(y=0)$ is
\begin{eqnarray}
u_i=\frac{1}{L}\frac{J_0(2\nu_i y/3)}{J_0(2\nu_i/\sqrt{15})},
\end{eqnarray}
where $J_n$ is the Bessel function of order $n$.
The normalization was determined according to~(\ref{norm}).
The Kaluza-Klein mass spectrum can be calculated from the boundary condition at
$z=2/5^{1/3}\;(y=\sqrt{3/5})$:
\begin{eqnarray}
\left.\frac{du_i}{dz}\right|_{z=(8/5)^{1/3}}\propto\;
\nu_iJ_1(2\nu_i/\sqrt{15})=0.
\end{eqnarray}
Thus we find $\nu_1 \simeq 7.42$, $\nu_2 \simeq 13.6$, $\nu_3 \simeq 19.7, \cdots$.
The Kaluza-Klein masses measured by an observer on the ring are
$\nu_i \ell^{-1}_0(\xi_0/\xi_*)\simeq0.855\times\nu_i\ell_0^{-1}$.

\begin{figure}[t]
  \begin{center}
    \includegraphics[keepaspectratio=true,height=50mm]{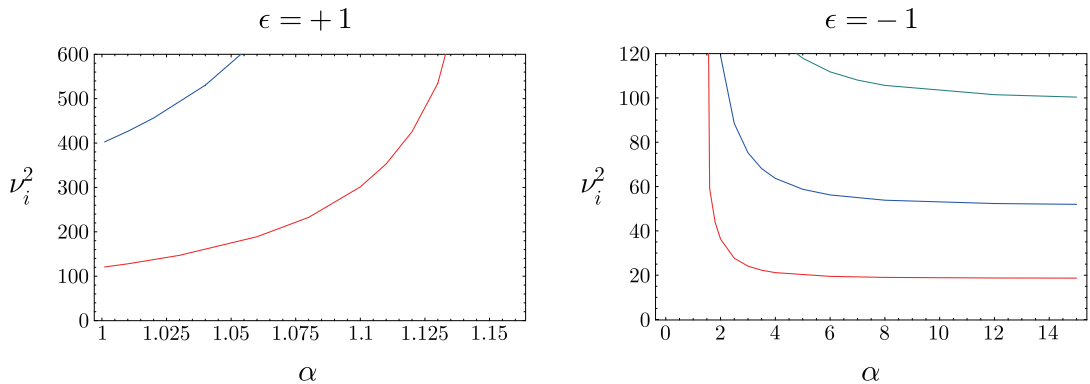}
  \end{center}
  \caption{The first several Kaluza-Klein eigenvalues as a function of $\alpha$.}%
  \label{fig:kkmass.eps}
\end{figure}

In the case of $\epsilon\neq 0$ we compute the mass spectra fully numerically.
The result is shown in figure~\ref{fig:kkmass.eps}.
As before, the Kaluza-Klein masses measured by an observer on the ring are
$\nu_i \ell^{-1}\alpha^{-1}$.
We are considering the case with $\alpha\sim O(1)$, and so
we have $m_i/\xi_* \gtrsim \ell^{-1}$.


\section{Discussion}\label{sec:final}

We have considered a warped braneworld in six dimensions.
The background is given by the model of~\cite{bolts} with a slight modification,
in which our universe is assumed to be a 4-brane wrapped around the axisymmetric
internal space.
Since the codimension of the brane is one,
this construction allows for localized matter on the brane.
We have performed a linearized perturbation analysis
(neglecting azimuthal excitations)
in order to study the long-distance behavior of weak
gravity sourced by arbitrary matter on the brane.
We have found that there are two scalar modes, $\zeta$ and $\Upsilon$, relevant to brane gravity.
The first one, $\zeta$, describes the shift of the brane position
and plays an important role in recovering the tensor structure of 4D gravity,
as in the 5D Randall-Sundrum construction~\cite{G-T}.
The mode $\Upsilon$ encodes the fluctuation of the volume of the internal space
(or that of the circumference of the 4-brane)
and signals a scalar-tensor theory of gravity.
However, the effect of $\Upsilon$ was shown to be
suppressed on scales greater than $\ell$ (or $\ell_0$).
Discrete Kaluza-Klein modes are Yukawa-suppressed at long distances.
Thus, we have successfully obtained standard 4D gravity on the brane.

The hybrid braneworld does not eliminate the hierarchy problem
with relatively ``large'' extra dimensions,
because one of the extra dimensions will be quite small compared to the other.
Indeed, the relation~(\ref{newton_const}) can be rewritten conveniently as
\begin{eqnarray}
M_{{\rm Pl}}^2=(M_6^4)\ell{\cal C}\;\frac{2(\xi_*^3-\xi_0^3)}{3\xi_*^2\sqrt{f_*}}\sim
(M_6^4)\ell{\cal C},\label{pl=6}
\end{eqnarray}
where $M_{{\rm Pl}}^2=\kappa_4^{-2}$ and $M_6^4=\kappa^{-2}$.
(For $\epsilon=0$, Eq.~(\ref{pl=6}) should be $M_{{\rm Pl}}^2=2(M_6^4)\ell_0{\cal C}/\sqrt{15}$.)
The circumference of the ring must be ${\cal C}\lesssim 10^{-16}$\;cm.
Thus, for $\ell\lesssim 10^{-2}$\;cm we get the fundamental scale $M_6\gtrsim 10^7$\;GeV.

We can easily configure
the present model with 4D de Sitter geometry~\cite{bolts}.
Constructing a Friedmann-Robertson-Walker
braneworld will also be possible by considering a moving brane in a warped bulk~\cite{branecos}
(see, however, the recent work of~\cite{reg_cos}).
It would be interesting to explore further various aspects of hybrid braneworlds.

\acknowledgments

We wish to thank Masato Minamitsuji for comments on the manuscript.
TK and YT are supported by the JSPS under Contract Nos.~19-4199 and 17-53192.


\appendix

\section{Gauge transformations}\label{App:gauge}

Under an infinitesimal coordinate transformation,
$x^{\mu}\to x^{\mu}+\delta x^{,\mu}$, $\xi\to\xi+\delta\xi$, and $\theta\to\theta+\delta\theta$,
the metric perturbations transform as
\begin{eqnarray}
&&\Psi\to\Psi-\frac{1}{\xi}\delta\xi,\qquad E\to E-\delta x,
\qquad
B\to B-\frac{\ell^2}{f}\delta\xi-\xi^2\delta x',
\qquad D\to D-\beta^2\ell^2f \delta\theta,
\nonumber\\
&&\Xi\to\Xi-\delta\xi'+\frac{f'}{2f}\delta\xi,
\qquad T\to T-\delta\theta',
\qquad
\Omega\to\Omega+\frac{f'}{2f}\delta\xi+\frac{3}{\xi}\delta\xi,
\label{gt}
\end{eqnarray}
and the perturbed gauge potential transforms as
\begin{eqnarray}
\delta A\to\delta A- A_{\theta}\delta\theta,
\qquad
\delta A_{\xi}\to\delta A_{\xi}-A_{\theta}\delta\theta',
\qquad
\delta A_{\theta}\to\delta A_{\theta}-A_{\theta}'\delta\xi.
\end{eqnarray}
To solve the 6D field equations it is convenient to choose
the gauge in which $E=B=T=0$. This is an analogue to the longitudinal gauge.
In this gauge, the position of the brane is also perturbed and is given by $\xi_*+\zeta(x)$.

The above gauge is in a sense ``bulk-based,''
and it will be more convenient to use the Gaussian-normal gauge
(i.e., ``brane-based'' coordinates) when looking at the Israel conditions on the brane.
The Gaussian-normal gauge is defined by $\overline{\delta g}_{\xi M}=0$,
where we use a bar to denote perturbations in this gauge.
We also impose that the position of the brane is not perturbed.
Then, from~(\ref{gt}) one finds that
the two gauges are related by a gauge transformation $\bar x^M\to x^M+\delta x^M$
such that
\begin{eqnarray}
\frac{\ell^2}{f}\delta\xi+\xi^2\delta x'=0,
\qquad
\Xi-\delta\xi'+\frac{f'}{2f}\delta\xi=0,
\qquad
\delta\theta'=0,
\end{eqnarray}
and
\begin{eqnarray}
\zeta+\delta\xi|_{\xi_*}=0.
\end{eqnarray}
We can fix the residual gauge freedom by imposing $\delta x|_{\xi_*}=\delta\theta|_{\xi_*}=0$.

The metric perturbations and gauge field induced on the brane are given by
$\overline{\delta g}_{ab}|_{\xi_*}$ and $\overline{\delta A}_{a}|_{\xi_*}$.
Hence, we have, for example,
\begin{eqnarray}
\overline{\Psi}_* = \Psi_*+\frac{1}{\xi_*}\zeta 
\qquad\text{and}\qquad
\overline{\delta A}_{\theta*}=\delta A_{\theta*}+A_{\theta*}'\zeta. 
\label{induced_pert}
\end{eqnarray}

\section{Vector modes}

Here we briefly summarize the properties of vector modes~\cite{Yoshiguchi, extended, Kinoshita}.
The vector perturbations are
\begin{eqnarray}
\delta g_{\mu\nu}=2\xi^2E_{(\mu,\nu)},\qquad
\delta g_{\mu\xi}=B_{\mu},\qquad
\delta g_{\mu\theta}=D_{\mu},
\end{eqnarray}
and $\delta A_{\mu}$, where $E_{\mu}^{\; ,\mu}=0, \cdots$.
Under a vector gauge transformation, $x^{\mu}\to x^{\mu}+\delta x^{\mu}$,
the variables transform as
\begin{eqnarray}
E_{\mu}\to E_{\mu}-\delta x_{\mu},\qquad
B_{\mu}\to B_{\mu}-\xi^2\delta x_{\mu}', \label{gtvec}
\end{eqnarray}
while $D_{\mu}$ and $\delta A_{\mu}$ are invariant.

From~(\ref{gtvec}) we find a gauge invariant combination
\begin{eqnarray}
V_{\mu}:=B_{\mu}-\xi^2 E_{\mu}'.
\end{eqnarray}
The Einstein equations read
\begin{eqnarray}
(\mu\nu):\;&&V_{\mu}'+\frac{2}{\xi}V_{\mu}+\frac{f'}{f}V_{\mu}=0,
\label{veq1}\\
(\mu\xi):\;&&\Box V_{\mu}=0.
\label{veq2}
\end{eqnarray}
Eq.~(\ref{veq2}) implies that only the zero mode is present for $V_{\mu}$.
Eq.~(\ref{veq1}) is then solved to give $V_{\mu}=\xi^{-2}f^{-1}c_{\mu}(x)$,
where $c_{\mu}$ is an integration constant.
However, the regularity at $\xi=\xi_0$ requires $c_{\mu}=0$.

The $(\mu\theta)$ component of the Einstein equations and
the $\mu$ component of the Maxwell equations yield the coupled equations of
motion for $D_{\mu}$ and $\delta A_{\mu}$.
These modes are not particularly interesting because they do not couple to
matter on the brane, $T_{\mu\nu}$ and $T_{\theta\theta}$, via the junction conditions.




\end{document}